\documentclass[global,twocolumn]{svjour}

\usepackage{makeidx}
\usepackage{amssymb}
\usepackage[sort&compress,square,comma]{natbib}
\usepackage[intlimits]{amsmath}
\usepackage{graphicx}
\usepackage[cp1250]{inputenc}
\usepackage[ps2pdf,hyperindex=true,final=true,colorlinks=true,pagebackref=false,bookmarks=true,bookmarksopen=true,bookmarksnumbered=true]{hyperref}

\newcommand{\Cu}{\ensuremath{\mathrm{Cu}}}
\newcommand{\Ni}{\ensuremath{\mathrm{Ni}}}
\newcommand{\Nb}{\ensuremath{\mathrm{Nb}}}
\newcommand{\Al}{\ensuremath{\mathrm{Al}}}
\newcommand{\Si}{\ensuremath{\mathrm{Si}}}

\newcommand{\Ar}{\ensuremath{\mathrm{Ar}}}
\renewcommand{\O}{\ensuremath{\mathrm{O}}}

\journalname{Applied Physics A}
\begin{document}
\title{Ferromagnetic $0$--$\pi$ Josephson junctions}
\author{M. Weides \inst{1} \and H. Kohlstedt \inst{1} \and R. Waser \inst{1} \and  M. Kemmler \inst{2} \and J. Pfeiffer \inst{2}\and D. Koelle \inst{2} \and R. Kleiner \inst {2} \and E. Goldobin \inst{2}}

\institute{Center of Nanoelectronic Systems for Information Technology (CNI), Research Centre J\"ulich, D-52425 J\"ulich, Germany \and Physikalisches Institut-Experimentalphysik II, Universit\"at T\"ubingen, Auf der Morgenstelle 14, D-72076 T\"ubingen, Germany}
\date{Received: date / Revised version: date}
%
\maketitle
\begin{abstract}
We present a study on low-$T_c$ superconductor-insulator-ferromagnet-superconductor (SIFS) Josephson junctions. SIFS junctions have gained considerable interest in recent years because they show a number of interesting properties for future classical and quantum computing devices.
We optimized the fabrication process of these junctions to achieve a homogeneous current transport, ending up with high-quality samples. Depending on the thickness of the ferromagnetic layer and on temperature, the SIFS junctions are in the ground state with a phase drop either $0$ or $\pi$. By using a ferromagnetic layer with variable step-like thickness along the junction, we obtained a so-called $0$-$\pi$ Josephson junction, in which $0$ and $\pi$ ground states compete with each other. At a certain temperature the $0$ and $\pi$ parts of the junction are perfectly symmetric, i.e. the absolute critical current densities are equal. In this case the degenerate ground state corresponds to a vortex of supercurrent circulating clock- or counterclockwise and creating a magnetic flux which carries a fraction of the magnetic flux quantum $\Phi_0$.
\end{abstract}
\section{Introduction}
Superconductivity (S) and ferromagnetism (F) are two competing phenomena. On one hand a bulk superconductor expels the magnetic field (Meissner effect). On the other hand the magnetic field for $H>H_{c2}$ destroys the superconductivity. This fact is due to the unequal symmetry in time: ferromagnetic order breaks the time-reversal symmetry, whereas conventional superconductivity relies on the pairing of time-reversed states. It turns out that the combination of both, superconductor and ferromagnet, leads to rich and interesting physics. One particular example -- the phase oscillations of the superconducting Ginzburg-Landau order parameter inside the ferromagnet -- will play a major role for the devices discussed in this work.\par
The current-phase relation $I_s(\phi)$ of a conventional SIS Josephson junction (JJ) is given by $I_s(\phi)=I_c \sin(\phi)$. $\phi=\theta_1-\theta_2$ is the phase difference of the
macroscopic superconducting wave functions $\Psi_{1,2}=\sqrt{n_s}e^{i\theta_{1,2}}$ (order-parameters of each electrode) across the junction, $I_c$ is the critical current. Usually $I_c$ is positive and the minimum of the Josephson energy $U=E_J(1-\cos{\phi})$, $E_J=\frac{I_c\Phi_0}{2\pi}$ is at $\phi=0$. However, Bulaevski\u{i} \emph{et al.} \cite{Bulaevskii1977} calculated the supercurrent through a JJ with ferromagnetic impurities in the tunnel barrier and predicted a negative supercurrent, $I_c<0$. For $-I_c\sin{(\phi)}=0$ the solution $\phi=0$ is unstable and corresponds to the maximum energy $U=E_J(1+\cos{\phi})$, while $\phi=\pi$ is stable and corresponds to the ground state. Such JJs with $\phi=\pi$ in ground state are called $\pi$ junctions, in contrast to conventional $0$ junctions with $\phi=0$. In case of a $\pi$ Josephson junction the first Josephson relation is modified to $I_s(\phi)=-I_c \sin(\phi)=I_c \sin(\phi +\pi)$. In experiment the measured critical current in a single junction is always positive and is equal to $|I_c|$. It is not possible to distinguish $0$ JJs from $\pi$ JJs from the current-voltage characteristic (IVC) of a single junction. The particular $I_c(T)$ \cite{Ryazanov2000} and $I_c(d_F)$ \cite{Kontos02Negativecoupling} dependencies for SFS/SIFS type junction are used to determine the $\pi$ coupled state. For low-transparency SIFS junctions the $I_c(d_F)$ dependence is given by
\begin{equation}\displaystyle I_c(d_F)\propto \exp{\left(\frac{-d_F}{\xi_{F1}}\right)}\;
  \cos{\left(\frac{d_F-d_F^{\mathrm{dead}}}{\xi_{F2}}\right)}\label{Eq:IcdF},\end{equation}
where $\xi_{F1},\xi_{F2}$ are the decay and oscillation lengths of critical current and $d_F^\mathrm{dead}$ is the dead magnetic layer thickness \cite{WeidesHighQualityJJ}. For $\frac{1}{2}\xi_{F2}\pi<d_F-d_F^{\mathrm{dead}}<\frac{3}{2}\xi_{F2}\pi$ the coupling in ground state of JJs is shifted by $\pi$.\par
\begin{figure}[b]
\resizebox{8.4cm}{!}{
  \includegraphics{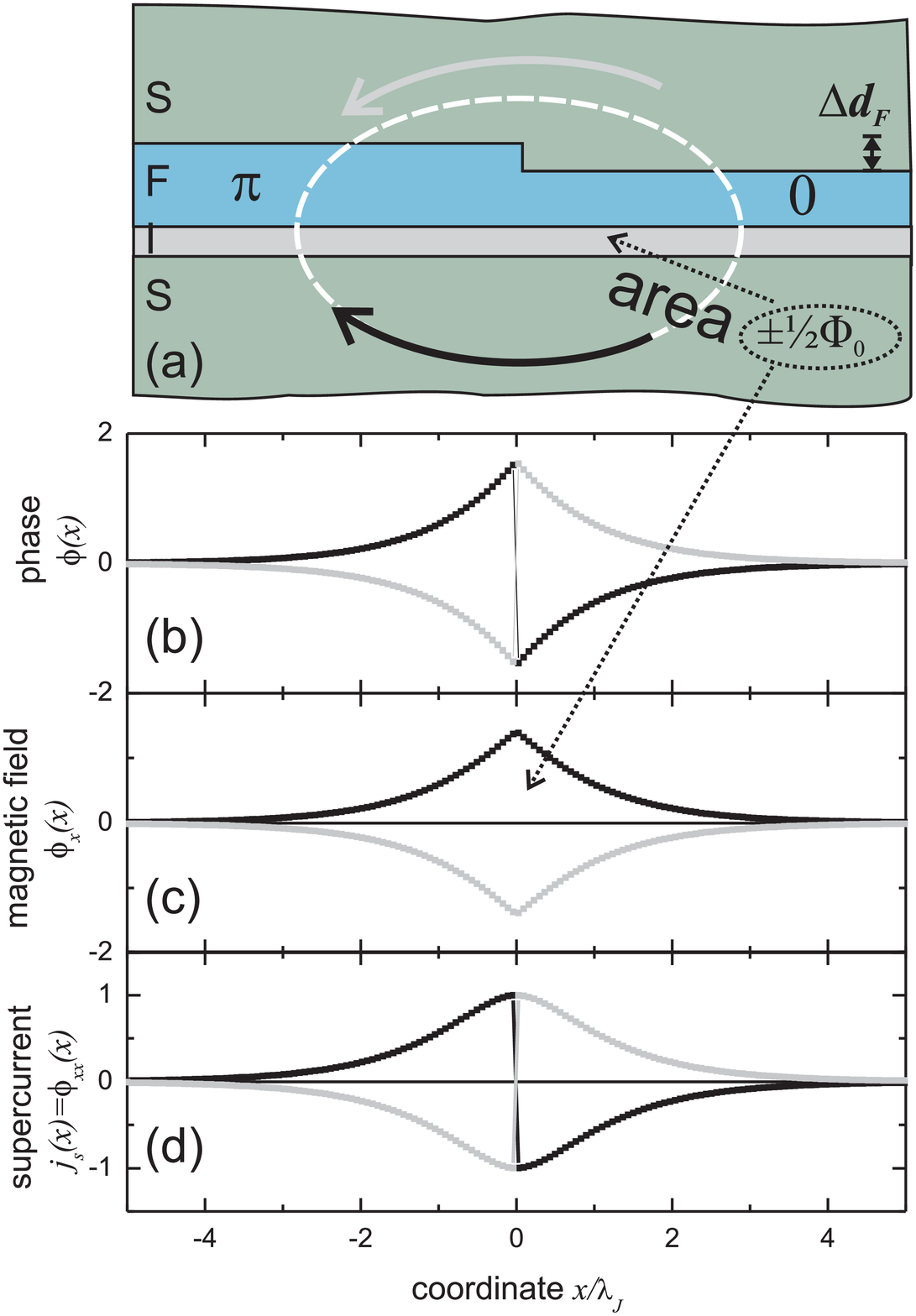}}
  \caption{(a) Sketch of a $0$--$\pi$ SIFS JJ with step-like thickness of F-layer and circulating supercurrent $j_s$ around $0$--$\pi$ phase boundary. The junction length $L\gg\lambda_J$, therefore the spontaneous flux (area below magnetic field) is equal to half of a flux quantum $\Phi_0$ (semifluxon). (b)-(d) depicts the phase $\phi(x)$, magnetic field $\phi_x(x)$ and supercurrent $j_s(x)=\frac{I_c}{|I_c|}\sin\phi$ of the $0$--$\pi$ junction.}
  \label{Sketch_0piSIFS}
\end{figure}

In a second work Bulaevski\u{i} \emph{et al.}  \cite{Bulaevskii1978} predicted the appearance of a \emph{spontaneous} supercurrent at the boundary between a $0$ and a $\pi$ coupled long JJ (LJJ). This supercurrent emerges in the absence of a driving bias current or an external field $H$, i.e. in the ground state. Depending on the length of the junction $L$ the supercurrent carries one half of the flux quantum, i.e. $\Phi_0/2$ (called \em semifluxon\em), or less. Fig.~\ref{Sketch_0piSIFS}(a) depicts the cross section of a symmetric $0$--$\pi$ {\it long} JJ. The spontaneous supercurrent $j_s$ flows either clockwise or counterclockwise, creating the magnetic field of $\pm\Phi_0/2$. The current density jumps from maximum positive to maximum negative value at the $0$--$\pi$ phase boundary. A theoretical analysis based on the perturbed sine-Gordon equation is given in Ref.~\cite{Goldobin02SF}. Below we will first discuss the properties of the spontaneous supercurrent and, second, various systems having $0$--$\pi$ phase boundaries.\par

\paragraph{Spontaneous supercurrent}
Kirtley \emph{et al.} \cite{Kirtley:IcH-PiLJJ} calculated the free energy of $0$--$\pi$ JJs for various
lengths of the $0$ and $\pi$ parts as a function of the normalized length $\ell=L/\lambda_J$ and the degree of asymmetry $\Delta= |j_c^\pi|L_{\pi}/|j_c^0|L_0$, where $j_c^{0},j_c^{\pi}$ are the critical current densities and $L_0,L_{\pi}$ are the lengths of $0$ and $\pi$ parts respectively, so that $L=L_0+L_{\pi}$. The state of a \em symmetric \em $0$--$\pi$ junction ($\Delta=1$) with spontaneous flux has lower energy than the states $\phi=0$ or $\phi=\pi$ without flux. Symmetric $0$--$\pi$ junctions have \emph{always} some self-generated spontaneous flux, although its amplitude vanishes for $L\rightarrow0$ as $\Phi\approx \Phi_0\ell^2/8\pi$. For example, a symmetric $0$--$\pi$ JJ of the total length $L=\lambda_J$ has a spontaneous magnetic flux $\Phi\approx0.04\Phi_0$ and a symmetric $0$--$\pi$ JJ with $L=8\lambda_J$ has a spontaneous flux of some $2-3\%$ below $\Phi_0/2$. Only in case of a infinitely long JJ we refer to the spontaneous flux as \emph{semifluxons}, for shorter JJs it is named \emph{fractional vortex}.\\
The supercurrent or magnetic flux can be directly detected by measuring $I_c(H)$ \cite{Kirtley:IcH-PiLJJ}, by scanning SQUID (superconducting quantum interference device) microscopy (in the LJJ limit, see \cite{KirtleyImaging96PRL,Hilgenkamp:zigzag:SF}) or by LTSEM (low temperature scanning electron microscopy) \cite{GrossKoelle}.

\paragraph{$0$--$\pi$ junctions technology}
$0$--$\pi$ Josephson junctions with a spontaneous flux in the ground state are realized with various technologies. The presence of fractional vortex has
been demonstrated experimentally in $d$-wave superconductor based ramp zigzag junctions
\cite{Hilgenkamp:zigzag:SF}, in long Josephson $0$--$\pi$ junctions fabricated
using the conventional $\Nb/\-\Al\mbox{-}\Al_2\O_3/\Nb$ technology with a pair of
current injectors \cite{Goldobin04PRLDynamicsSF}, in the so-called
tricrystal grain-boundary LJJs \cite{Kirtley:SF:T-dep,KirtleyImaging96PRL,Sugimoto:TriCrystal:SF} or in SFS/SIFS JJs \cite{FrolovRyazanovSemifluxon,RoccaAprili05classicalspins,WeidesFractVortex} with \emph{stepped} ferromagnetic barrier as in Fig.~\ref{Sketch_0piSIFS}. In the latter systems the Josephson phase in the ground state is set to $0$ or $\pi$ by choosing proper F-layer thicknesses $d_1$, $d_2$ for $0$ and $\pi$ parts, i.e. the amplitude of the critical current densities $j^0_c$ and $j^\pi_c$ can be controlled to some degree. The advantages of this
system are that it can be prepared in a multilayer geometry (allowing topological flexibility) and it can be easily combined with the
well-developed $\Nb/\Al\mbox{-}\Al_2\O_3/\Nb$ technology.\\
The starting point for estimation of the ground state of a \emph{stepped} JJ is studying the IVCs and $I_c(H)$ for the \emph{planar} reference $0$ and $\pi$ JJs. From this one can calculate important parameters such as the critical current densities $j_c^0$, $j_c^\pi$, the
Josephson penetration depths $\lambda_J^0$, $\lambda_J^\pi$ and the ratio of asymmetry
$\Delta$. For $0$--$\pi$ junctions one
needs $0$ and $\pi$ coupling in \emph{one} junction, setting high demands on
the fabrication process. The ideal $0$--$\pi$ JJ would have equal
$|j_c^0|=|j_c^\pi|$ and a $0$--$\pi$ phase boundary in its center to have a symmetric situation. Furthermore the junctions should be underdamped (SIFS
structure) since low dissipation is necessary to study dynamics and eventually macroscopic quantum effects. The junctions should have a high $j_c$ (and hence small $\lambda_J\propto\sqrt{j_c}$) to reach the LJJ
limit and to keep high $V_c=I_cR$ products, where $V_c$ is the characteristic voltage and $R$ the normal state resistance.\\
Previous experimental works on $0$--$\pi$ JJs based on SFS technology
\cite{FrolovRyazanovSemifluxon,RoccaAprili05classicalspins} gave no
information about $j^0_c$ and $j^\pi_c$. Hence, the Josephson penetration
depth $\lambda_J$ could not be calculated for these samples and the ratio of
asymmetry $\Delta$ was unknown. The first intentionally made symmetric $0$--$\pi$ tunnel JJ of
SIFS type with a large $V_c$ was realized by the authors \cite{WeidesFractVortex}, making direct transport measurements of $I_c(H)$ and calculation of the ground state with spontaneous flux feasible.\par
Within this paper we \emph{review} the physics of $0$--$\pi$ coupled SIFS-type Josephson junctions and give an overview on our experimental results. Special focus is put on the fabrication of SIFS junctions having a planar or stepped-typed ferromagnetic layer ($\Ni\Cu$), the determination of ground state ($0$ or $\pi$ for planar JJs) and asymmetry of critical currents (stepped JJs). Finally we give an estimation of the spontaneous magnetic flux in the ferromagnetic $0$--$\pi$ JJs.

\section{Fabrication}
\begin{figure}[tb]
\resizebox{8.4cm}{!}{
\includegraphics{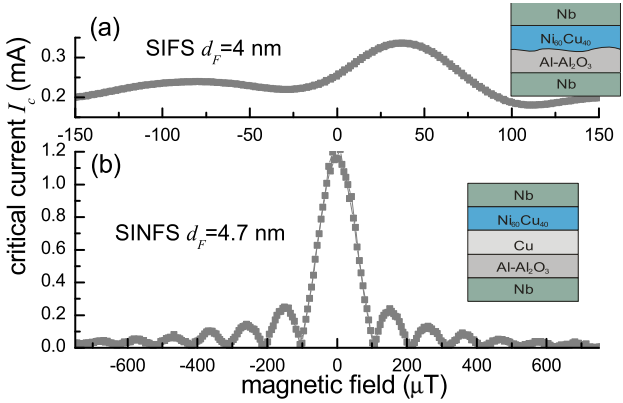}}
\caption{\label{Fig:IcH_SINFS}(Color online) $I_c(H)$ of (a) SIFS (4 nm $\Ni\Cu$) and (b) SINFS ($\Cu$ 2 nm, $\Ni\Cu$ 4.7 nm)
stacks. Oxygen pressure is 0.45 mbar for SIFS and 0.015 mbar for SINFS type.}
\end{figure}

The fabrication process for \emph{planar} junctions is based on $\Nb/\Al\mbox{-}\Al_2\O_3/\Ni\Cu/\Nb$ stacks, deposited by dc magnetron sputtering \cite{WeidesFabricationJJPhysicaC}. Thermally oxidized $4$-inch $\Si$ wafer served as substrate. First of all, a $120$ nm thick $\Nb$ bottom electrode and a $5$ nm thick $\Al$ layer were deposited. Second, the aluminium was oxidized for $30$ min at room temperature in a separate chamber. Third, the ferromagnet (i.e. $\Ni_{60}\Cu_{40}$ alloy, $T_C=225\;\rm{K}$) was deposited. To have many structures with different thicknesses in one fabrication run, we decided to deposit a \emph{wedge}-shaped F-layer. For this the substrate and sputter target were shifted about half of the substrate diameter. This allowed the preparation of SIFS junctions with a gradient in F-layer thickness in order to minimize inevitable run-to-run variations. The sputtering rates for $\Ni\Cu$ along the gradient were determined by thickness measurements on reference samples using a Dektak profiler. At the end a $40$ nm $\Nb$ cap layer was deposited. The tunnel junctions were patterned using a three level optical  photolithographic mask procedure and $\Ar$ ion-beam milling \cite{Gurvitch82NbAlONb}. The insulation between top and bottom electrode is done by a self-aligned growth of $\Nb_2\O_5$ insulator by anodic oxidation of $\Nb$ after the ion-beam etching.  The $\Nb_2\O_5$ exhibited a defect free insulation between the superconducting electrodes.\\
Topological and electrical measurements, see Ref. \cite{WeidesFabricationJJPhysicaC}, indicated that the direct deposition of $\Ni\Cu$ on the tunnel barrier (SIFS-stacks) led to an anomalous $I_c(H)$ dependence such as shown in Fig. \ref{Fig:IcH_SINFS}(a), which is an indication for an inhomogeneous current transport. An additional $2\;\rm{nm}$ thin $\Cu$ layer between the $\Al_2\O_3$ tunnel barrier and the ferromagnetic $\Ni\Cu$  (SINFS-stacks) brought considerable benefits, as it ensured a homogeneous current transport, see Fig. \ref{Fig:IcH_SINFS}(b). In this way a high number of functioning devices with $j_c$ spreads less than 2\% was obtained. The variation of the F-layer thickness over a length of one junction diameter is less than $0.02\;\rm{nm}$. For simplification we refer in the following to SIFS stacks, although the actual multilayer is SINFS-type. \\
The patterning of \emph{stepped} junctions was done after the complete
deposition of the planar SIFS stack and before the definition of the junction mesa by
argon-etching and $\Nb_2\O_5$ insulation. The detailed process is published in Ref.~\cite{WeidesSteppedJJ}. The junction was partly
protected with photoresist to define the step location in the F-layer, followed by i) \emph{selective reactive etching} of the $\Nb$, ii) \emph{ion-etching} of the $\Ni\Cu$ by $\Delta d_F$ and iii) subsequent \emph{in situ} deposition of $\Nb$.
To our knowledge, this was the first controlled patterning of $0$--$\pi$ JJs based on a ferromagnetic interlayer.\\The planar $0$, $\pi$ reference junctions and the stepped $0$--$\pi$ junctions were fabricated from a single trilayer.

\section{SIFS junctions without step-like F-layer}
\begin{figure}[tb]
\resizebox{8.4cm}{!}{
\includegraphics{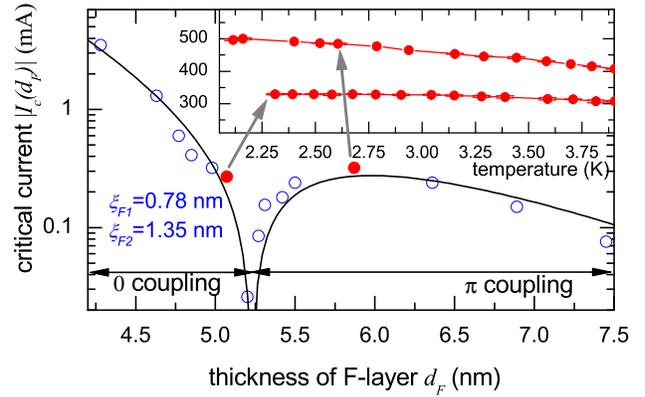}}
\caption{\label{Fig:IcdF}(Color online) $I_c(d_F)$ and $I_c(T)$ (inset) dependences of SIFS
junctions at $4.2\;\rm{K}$. Note the difference of the slope of $I_c(T)$ for $0$ and $\pi$ coupled junction (inset).}
\end{figure}

All investigated junctions had an area of $10\;000\:\mathrm{\mu m^2}$, but the length and width were different for different junctions. The length was comparable or shorter than the Josephson penetration depth $\lambda_J$. We investigated the thickness dependence of the critical current $I_c(d_F)$. To produce the
$\Al_2\O_3$ barrier the $\Al$ layer was oxidized at $0.015\;\rm{mbar}$ yielding $j_c \approx 4.0\:\mathrm{kA/cm^2}$ for the reference super\-conductor-insulator-superconductor (SIS) JJs. Then SIFS stacks with wedge-like F-layer were fabricated in another run. Taking the JJs of the same geometry ($100\times100\;\rm{\mu m^2}$), but situated at different places on the wafer (i.e. different $d_F$) we have measured the nonmonotonic $I_c(d_F)$ dependence shown in Fig.~\ref{Fig:IcdF}. As a result the fitted parameters are $\xi_{F1}=0.78\;\mathrm{nm}$, $\xi_{F2}=1.35\;\mathrm{nm}$ and $d_F^{\mathrm{dead}}\approx 3.09\;\mathrm{nm}$. The coupling changed from $0$ to $\pi$ at the crossover thickness $d_{F}^{0\mbox{-}\pi} = \frac{\pi}{2}\xi_{F2}+d_F^\mathrm{dead}=5.21\:\rm{nm}$ \cite{WeidesHighQualityJJ}.

The magnetic and spin-orbit scattering in the F-layer mixes the up and down spin states of electrons in the conduction bands. If the spin-flip scattering time $\tau_s$ is short $\hbar \tau_s^{-1}\gg k_B T_c$, like in $\Ni\Cu$ alloys, the temperature dependence of scattering provides the dominant mechanism for the $I_c(T)$ dependence \cite{OboznovRyazanov06IcdF}. The oscillation period $\xi_{F2}$ becomes shorter for decreasing temperature, thus the whole $I_c(d_F)$ dependence is squeezed to thinner F-layer thicknesses. Hence, the temperature dependence of the critical current $I_c(T)$ is an interplay between an increasing component due to an increasing gap and a magnetic coupling dependent contribution which may de- or increase $I_c$. The $I_c(T)$ relations for two JJs (one $0$, one $\pi$) are shown in the inset of Fig.~\ref{Fig:IcdF}. At $d_F=5.11\:\mathrm{nm}$ the JJ is $0$ coupled, but one can relate the nearly constant
$I_c$ below $3.5\:\mathrm{K}$ to the interplay between an increasing gap and a decreasing
oscillation length $\xi_{F2}(T)$. The $d_F=5.87\:\mathrm{nm}$ JJ is $\pi$ coupled and showed a linearly increasing $I_c$ with decreasing temperature.

\section{SIFS junctions with step-like F-layer}

\begin{figure}[tb]
\resizebox{8.4cm}{!}{
\includegraphics{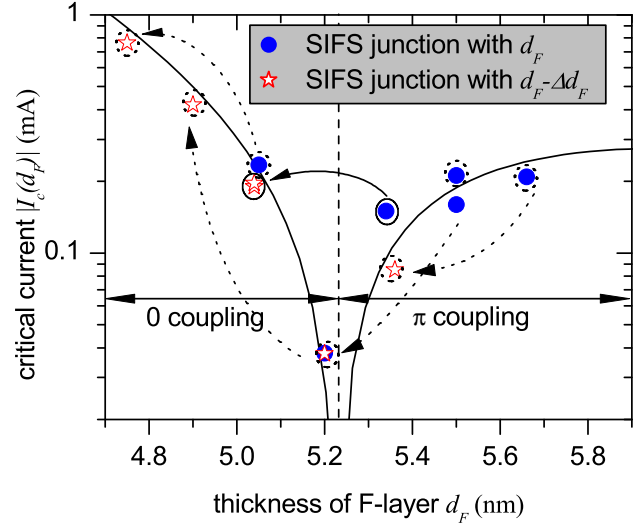}}
\caption{\label{Fig:IcdF_Etch}(Color online) Critical current $I_c$ of the uniformly etched (star) and non-etched (dot) SIFS junctions versus the F-layer thickness before etching $d_F$. The fit of the experimental data for non-etched samples using Eq.(\ref{Eq:IcdF}) is shown by the continuous line. The JJs were oxidized at $0.015\;\rm{mbar}$.}
\end{figure}

Various structures on the wafer were placed within a narrow ribbon perpendicular to the gradient in the F-layer thickness and were replicated along this gradient. One ribbon contained reference JJs with the uniform F-layer thickness $d_1$ (uniformly etched) and $d_2$ (non-etched) and a JJ with a step $\Delta d_F$ in the F-layer thickness from $d_1$ to $d_2$. The lengths $L_{d_1}$ and $L_{d_2}$ are both equal to $167\;\rm{\mu m}$. The lithographic accuracy is of the order of $1\;\rm{\mu m}$. A set of structures  with difference in $d_F$ between neighboring ribbons of $0.05\;\rm{nm}$ was obtained.
Comparing the critical currents $I_c$ of non-etched JJs (dots), see Fig.~\ref{Fig:IcdF_Etch} with the experimental $I_c(d_F)$ data for the etched samples (stars) we estimate the etched-away F-layer thickness as $\Delta d_F \approx 0.3\;\rm{nm}$. The stars in Fig.~\ref{Fig:IcdF_Etch} are shown already shifted by this amount. Now we choose the set of junctions which have the thickness $d_2$ and critical current $I_c(d_2)<0$ ($\pi$ junction) before etching and have the thickness $d_1=d_2-\Delta d_F$ and critical current $I_c(d_1) \approx -I_c(d_2)$ ($0$ junction) after etching. One option is to choose the junction set denoted by closed circles around the data points in Fig.~\ref{Fig:IcdF_Etch}, i.e. $d_1=5.05\;\rm{nm}$ and $d_2=5.33\;\rm{nm}$.\par

\begin{figure}[tb]
\resizebox{8.4cm}{!}{
\includegraphics{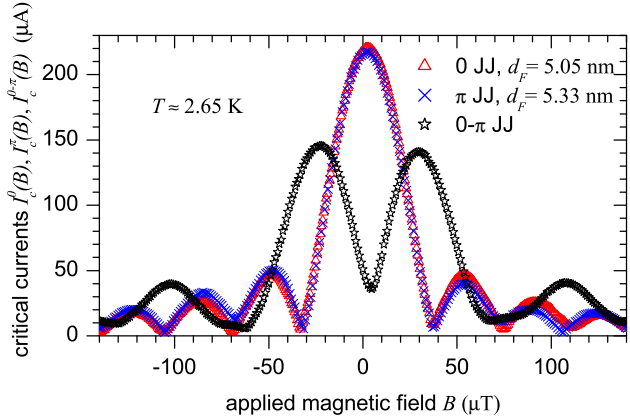}}
\caption{\label{Fig:IcH-0-pi}(Color online) $I_c(H)$ of $0$--$\pi$ JJ (open triangles) with $H$ applied parallel to short axis, overlayed with the non-etched (dot) and etched (stars) reference SIFS junction measurements. At $T \approx 2.65\;\rm{K}$ the $0$--$\pi$ JJ becomes symmetric. The junction dimensions are $330\times30\;\rm{\mu m^2}$.}
\end{figure}

The \emph{I-V} characteristics and the magnetic field dependence of the critical current $I_c(H)$ was measured for all three junctions: $0$ JJ with $d_F=d_1$, $\pi$ JJ with $d_F=d_2$ and $0$--$\pi$ JJs with stepped F-layer ($d_1$ and $d_2$ in each half). The magnetic diffraction pattern $I_c(H)$ of the $0$--$\pi$ JJ and the $0$ and $\pi$ reference JJs are plotted in Fig. \ref{Fig:IcH-0-pi}. The magnetic field $H$ was applied in-plane of the sample and parallel to the step in the F-layer. Due to a small net magnetization of the F-layers the $I_c(H)$ of references
junctions were slightly shifted along the $H$ axis. Nevertheless, both had the same oscillation period $\mu_0 H_{c1}\approx36\;\rm{\mu T}$. At $T\approx4.0\;\rm{K}$ the $0$--$\pi$ JJs was slightly asymmetric with $I_c^{0}\approx208\;\rm{\mu A}$ and $I_c^{\pi}\approx171\;\rm{\mu A}$ (data of reference JJs). To achieve a more symmetric configuration, the bath temperature was reduced, because a decrease in temperature should increase $I_c^\pi=I_c(d_2)$ more than $I_c^0=I_c(d_1)$, like for the $0$ and $\pi$ samples in the inset of Fig.~\ref{Fig:IcdF}. As a result, both $I_c^0(T)$ and $I_c^\pi(T)$ were increasing when decreasing the temperature, but with different rates. At $T\approx2.65\;\rm{K}$ the critical currents $I_c^{0}$ and $I_c^{\pi}$ became approximately equal, see Fig. \ref{Fig:IcH-0-pi}.  The magnetic field dependence of the planar reference junctions $I_c^{0}(H)$ and $I_c^{\pi}(H)$ look like perfect Fraunhofer patterns. One can see that the $I_c^0(H)$ and $I_c^\pi(H)$ measurements almost coincide, having the form of a symmetric Fraunhofer pattern with the critical currents $I_c^{0}\approx 220\;\rm{\mu A}$, $I_c^{\pi}\approx 217\;\rm{\mu A}$ and the same oscillation period. The stepped $0$--$\pi$ junction had a magnetic field dependence $I_c^{0\mbox{-}\pi}(H)$ with a clear minimum near zero field and almost no asymmetry. The critical currents at the left and right maxima ($146\;\rm{\mu A}$ and $141\;\rm{\mu A}$) differ by less than $4\;\rm{\%}$, i.e. the $0$--$\pi$ junction is symmetric, and its ground state in absence of a driving bias or magnetic field
($I=H=0$) can be calculated \cite{WeidesFractVortex}. Our symmetric $0$--$\pi$ LJJ had an normalized length of $\ell=1.3$, with a spontaneous flux in the ground state of\[ \pm\Phi\approx\Phi_0 \ell^2/8\pi\approx 0.067 \cdot\Phi_0,\]
being equal to $13\rm{\%}$ of $\Phi_0/2$. A detailed calculation taking several deviations from the ideal short JJ model into account can be found elsewhere \cite{WeidesPRB0Pi}.

\section{Summary}
The concept and realization of $0$-$\pi$ junction based on SIFS stacks has been presented. The realization of $\pi$ coupling in SIFS junctions and the precise combination of $0$ and $\pi$ coupled parts in a single junction has been shown. The coupling of the ferromagnetic Josephson tunnel junctions was investigated by means of transport measurements. The emergence of a spontaneous flux, which was calculated as $13\%$ of half a flux quantum $\Phi_0/2$, was observed in the magnetic field dependence of the current-voltage characteristics of the $0$--$\pi$ JJ.\\
As an outlook, the ferromagnetic $0$--$\pi$ Josephson junctions allow to study the physics of fractional vortices with a good temperature control of the symmetry between $0$ and $\pi$ parts. We note that symmetry is only needed for JJ lengths $L\lesssim\lambda_J$. For longer JJs the semifluxon appears even in rather asymmetric JJs, and $T$ can be varied in a wide range affecting the semifluxon properties only weakly. The presented SIFS technology allows us to construct $0$, $\pi$ and $0$--$\pi$ JJs with comparable $j_c^0$ and $j_c^{\pi}$ in a single fabrication run. Such JJs may be used to construct classical and quantum devices such as oscillators, memory cells, $\pi$ flux qubits \cite{Maekawa1,Maekawa2} or semifluxon based qubits \cite{Goldobin:2005:QuTu2Semifluxons}.
\section*{Acknowledgment}
This work was suportted by Heraeus Foundation and the Deutsche Forschungsgemeinschaft (projects SFB/TR 21).

\end{document}